\documentclass{article}
\usepackage{amsmath}
\usepackage{graphicx}
\usepackage{tabularx}

\input{tcilatex}

\begin{document}

\author{Salman Khan\thanks{%
sksafi@comsats.edu.pk} \\
Department of Physics, Comsats Institute of Information \\
Technology, Chak Shahzad, Islamabad, Pakistan.}
\title{Tripartite entanglement of fermionic system in accelerated frames}
\maketitle

\begin{abstract}
The dynamics of tripartite entanglement of fermionic system in noninertial
frames through linear contraction criterion when one or two observers are
accelerated is investigated. In one observer accelerated case the
entanglement measurement is not invariant with respect to the partial
realignment of different subsystems and for two observers accelerated case
it is invariant. It is shown that the acceleration of the frame does not
generate entanglement in any bipartite subsystems. Unlike the bipartite
states, the genuine tripartite entanglement does not completely vanish in
both one observer accelerated and two observers accelerated cases even in
the limit of infinite acceleration. The degradation of tripartite
entanglement is fast when two observers are accelerated than when one
observer is accelerated. It is shown that tripartite entanglement is a
better resource for quantum information processing than the bipartite
entanglement in noninertial frames .\newline
PACS: 03.65.Ud; 03.67.Mn;04.70.Dy

Keywords: Tripartite entanglement; Noninertial frames.
\end{abstract}

\section{Introduction}

Entanglement is not only one of the most striking properties of quantum
mechanics but also the essential tool for the practical realization of
quantum information and quantum computation \cite{springer}. The concepts of
all the subfields of quantum information theory, such as teleportation of
unknown states \cite{Bennett}, quantum key distribution \cite{Ekert},
quantum cryptography \cite{Bennett2} and quantum computation \cite{Grover,
Vincenzo}, are based on prior quantum entanglement between subsystems of a
composite system. The dynamics of entanglement in inertial frames of various
bipartite qubit and qutrit states have been extensively studied under
different conditions via different existing entanglement quantifiers. Since
multipartite states are as useful for quantum information processing as
bipartite states \cite{Hans,Karlsson}, therefore, it is important to
thoroughly investigate the dynamics of entanglement of multipartite states.
Unlike bipartite systems, though many criteria for quantifying entanglement
in pure and mixed tripartite systems are proposed \cite%
{Horodocki,Rudolph,Kia,Fabrizo,Karoly}, however, there is no single well
defined criterion for quantifying the entanglement of multipartite systems
and systems of higher dimensions to provide the necessary and sufficient
condition.

The investigations of entanglement dynamics in inertial frames show that the
total entanglement between various uniformly moving parties is conserved,
however, it may transfers from one set of degrees of freedom to others \cite%
{Many}. For a complete understanding of the behavior of entanglement between
various parties, it is necessary to investigate its dynamics in the
relativistic setup as well. Because the relativistic framework provides a
more complete picture and is important both from theoretical and
experimental perspective. Recently, the study of entanglement of various
fields in the accelerated frames has taken into account and valuable results
about the behavior of entanglement have been obtained \cite%
{Alsing,Ling,Gingrich,Pan, Schuller, Terashima, Muxin, Olson}. The effects
of noise on dynamics of entanglement under different quantum channels in
noninertial frames have also been studied \cite{Wang, Salman,Salman1,Salman2}%
. It is shown that for most channels the loss of entanglement is quicker and
entanglement sudden death may occur. On the other hand, the sudden rebirth
of entanglement is also reported under the action of some particular
channels. However, these studies are limited only to bipartite qubit systems
in the accelerated frames. More recently, the dynamics of entanglement in
fermionic and bosonic tripartite qubit systems in noninertial frames are
studied in Refs. \cite{Hwang, Wang2, Shamirzai} using $\pi $-tangle and
logarithmic negativity as the measurement of entanglement. These studies
show that the degree of entanglement is degraded by the acceleration of the
frames and, like the two-tangles in inertial frames for \textit{GHZ} state,
the two-tangles for \textit{GHZ} state are zero when one or two observers
are in the accelerated frames. It is also shown that with increasing
acceleration the degradation of entanglement is slower in the case of
tripartite states as compared to bipartite states in noninertial frames.
Unlike bipartite entanglement in noninertial frames, the tripartite
entanglement does not completely vanish even in the limit of infinite
acceleration.

In this paper we investigate the effect of acceleration of noninertial
frames on both bipartite and tripartite entanglement of Dirac field using
partial realignment criterion (linear contraction) \cite{Horodocki} as
entanglement quantifier. We consider the Dirac fields as shown in Refs. \cite%
{AsPach, Martin, Bruschi}. Our system consists of three observers; Alice,
Bob and Charlie. First we consider only one observer (Charlie) in the
accelerated frame and then we do calculations for two accelerated observers
(Bob and Charlie). We show that in either case the acceleration does not
produce entanglement in any of the bipartite subsystems and the genuine
tripartite entanglement degrades with increasing acceleration. We also show
that acceleration of the frame affects the uniform distribution of the three
way entanglement among the parties.

This paper is organized as follows. In section $2$ we explain our system and
derive the mixed density matrix whose entanglement dynamics are to be
studied. In section $3$ we briefly review the various criterion for the
measurement of tripartite entanglement. In particular, we briefly review the
three tangle, the $\pi $-tangle, realignment criterion and linear
contraction. In section $4$ we present our results. The last section
summarize our results.

\section{Tripartite state in accelerated frames}

To investigate the effect of acceleration on the dynamics of tripartite
entanglement, we consider three entangled fermionic modes whose frequencies
in an inertial frame are $\omega _{a}$, $\omega _{b}$, $\omega _{c}$ with
all other modes of the field in the vacuum state. The three observers are
provided with detectors each sensitive to a single mode such that Alice ($A$%
) records the particles with mode $\omega _{a}$, Bob ($B$) detector is
sensitive to mode $\omega _{b}$ and the detector of Charlie ($C$) is tuned
for the mode $\omega _{c}$. We will consider that either one observer
(Charlie) or two observers (Bob and Charlie) move with some constant
acceleration and the third observer (Alice) stays stationary. The field
equation can be solved in Minkowski coordinates, which are appropriate for
inertial observers and in Rindler coordinates for accelerated observers. To
cover Minkowski space, two different sets of Rindler coordinates that differ
from each other by an overall change in sign and define two causally
disconnected Rindler regions ($I,II$) are necessary (for detail see \cite%
{Alsing} and references therein). From the perspective of inertial observer,
the Dirac fields as shown in Refs. \cite{AsPach, Martin, Bruschi}, describe
a superposition of Minkowski monochromatic modes $|0\rangle _{M}=\otimes
_{i}|0_{\omega _{i}}\rangle _{M}$ and $|1\rangle _{M}=\otimes _{i}|1_{\omega
_{i}}\rangle _{M}$ $\forall i$ with%
\begin{align}
|0_{\omega _{i}}\rangle _{M}& =\cos r_{i}|0_{\omega _{i}}\rangle
_{I}|0_{\omega _{i}}\rangle _{II}+\sin r_{i}|1_{\omega _{i}}\rangle
_{I}|1_{\omega _{i}}\rangle _{II},  \notag \\
|1_{\omega _{i}}\rangle _{M}& =|1_{\omega _{i}}\rangle _{I}|0_{\omega
_{i}}\rangle _{II},  \label{1}
\end{align}%
where $r_{i}$ ($0\leq r_{i}\leq \pi /4$) is a dimensionless acceleration
parameter of the accelerated observer and is given by $\cos r_{i}=\left(
e^{-2\pi \omega _{i}c/a}+1\right) ^{-1/2}$. The parameters $\omega _{i}$, $c$
and $a$, in the exponential stand, respectively, for the frequency of $i$th
mode, speed of light in vacuum and acceleration of the accelerated observer.
The parameter $r_{i}=0$ for $a=0$ and $r_{i}=\pi /4$ for $a=\infty $. In Eq.
(\ref{1}), the subscripts $I$ and $II$ of the kets represent the modes
decomposition in the two causally disconnected regions in Rindler spacetime.
That is, each Minkowski mode $\omega _{i}$ has a Rindler mode expansion
given by Eq. (\ref{1}). In other words, the acceleration causes the
information initially formed in region $I$ to leak into region $II$. Since a
uniformly accelerated observer in region $I$ has no access to field modes in
the causally disconnected region $II$ and vice versa. Therefore, the
observer must trace over the modes in inaccessible region, losing
information about the state, which essentially results in the detection of a
thermal state. This effect is called the Unruh effect \cite{Davies,Unruh}.

We consider the following maximally entangled \textit{GHZ} state%
\begin{equation}
|\psi \rangle _{ABC}=\frac{1}{\sqrt{2}}\left( |0_{\omega _{a}}0_{\omega
_{b}}0_{\omega _{c}}\rangle _{A,B,C}+|1_{\omega _{a}}1_{\omega
_{b}}1_{\omega _{c}}\rangle _{A,B,C}\right) ,  \label{2}
\end{equation}%
where $|\cdot \cdot \cdot \rangle _{A,B,C}=|\cdot \rangle _{A}|\cdot \rangle
_{B}|\cdot \rangle _{C}$\ and the three capital alphabets in the subscripts
of the kets represent the three observers. In Eq. (\ref{2}) $|0_{\omega
_{a(b,c)}}\rangle _{A(B,C)}$ and $|1_{\omega _{a(b,c)}}\rangle _{A(B,C)}$,
respectively, represent the Minkowski vacuum state and Minkowski excited
state. In fact, a given Minkowski mode of a particular frequency spreads
over all positive Rindler frequencies ($\omega _{i}/(a/c)$) that peaks about
the Minkowski frequency \cite{Martin, Bruschi,Takagi,Alsing2}. However, to
simplify our problem we consider a single mode only in the Rindler region $I$%
, an approximation which is valid provided the detectors with the observers
are highly monochromatic that are sensitive only to their respective modes
of the chosen angular frequencies. If Alice stays stationary and the other
two observers move with some constant accelerations, then substituting the
Rindler modes from Eq. (\ref{1}) for the Minkowski modes in Eq. (\ref{2})
for the two accelerated observers gives%
\begin{align}
|\psi \rangle _{ABC}& =\frac{1}{\sqrt{2}}(\cos r_{b}\cos r_{c}|00000\rangle
_{A,BI,BII,CI,CII}  \notag \\
& +\cos r_{b}\sin r_{c}|00011\rangle _{A,BI,BII,CI,CII}  \notag \\
& +\sin r_{b}\cos r_{c}|01100\rangle _{A,BI,BII,CI,CII}  \notag \\
& +\sin r_{b}\sin r_{c}|01111\rangle _{A,BI,BII,CI,CII}  \notag \\
& +|11010\rangle _{A,BI,BII,CI,CII}).  \label{3}
\end{align}%
where $r_{b}$ and $r_{c}$ are the accelerations of Bob and Charlie,
respectively. In order to be handy and present the relation in simple form,
we dropped the frequencies in the subscript of each entry of the kets. Since
the Rindler modes in region $II$ are inaccessible, tracing out over those
modes, that is, over third and fifth qubits, leave the following initial
mixed density matrix%
\begin{align}
\rho _{ABC}& =\frac{1}{2}[\cos ^{2}r_{b}\cos ^{2}r_{c}|000\rangle \langle
000|+\cos ^{2}r_{b}\sin ^{2}r_{c}|001\rangle \langle 001|  \notag \\
& +\sin ^{2}r_{b}\cos ^{2}r_{c}|010\rangle \langle 010|+\sin ^{2}r_{b}\sin
^{2}r_{c}|011\rangle \langle 011|  \notag \\
& +\cos r_{b}\cos r_{c}(|000\rangle \langle 111|+|111\rangle \langle
000|)+|111\rangle \langle 111|].  \label{4}
\end{align}%
Note that we have also dropped the subscript $I$ that indicates the Rindler
modes in region $I$. In the rest of the paper, all calculations correspond
to the Rindler modes in region $I$.

\section{Entanglement measurement of tripartite states}

In literature, various criterion for measuring the entanglement of
tripartite states are suggested. The most popular among them are the
residual three tangle \cite{Coffman} and $\pi $-tangle \cite{Fan,Vidal}.
Other measurements for tripartite entanglement include realignment criterion 
\cite{Rudolph,Kia} and linear contraction \cite{Horodocki}. The realignment
and linear contraction criteria are comparatively easy in calculation and
are strong criteria for entanglement measurement that detects entanglement
of states for which other criteria fail. However it does not detect the
entanglement of all states. In the following we briefly review some of these
criteria.

The three tangle, which is polynomial invariant \cite{Verstraete,Leifer}, is
a good measure for tripartite entanglement of mixed density matrices.
However, it needs an optimal decomposition of a mixed density matrix which,
in general, is a tough enough task except in a few rare cases \cite{Lohmayer}%
. On the other hand, the $\pi $-tangle for a tripartite state $|\psi \rangle
_{ABC}$ is given by%
\begin{equation}
\pi _{ABC}=\frac{1}{3}(\pi _{A}+\pi _{B}+\pi _{C}),  \label{5}
\end{equation}%
where $\pi _{A}$ is the residual entanglement and is given by%
\begin{equation}
\pi _{A}=\mathcal{N}_{A(BC)}^{2}-\mathcal{N}_{AB}^{2}-\mathcal{N}_{AC}^{2}.
\label{6}
\end{equation}%
The other residual tangles ($\pi _{B},\pi _{C}$) are defined in a similar
way. In Eq. (\ref{6}), $\mathcal{N}_{AB}(\mathcal{N}_{AC})$ is a two-tangle
and is given as the negativity of mixed density matrix $\rho
_{AB}=Tr_{C}|\psi \rangle _{ABC}\langle \psi |$ $(\rho _{AC}=Tr_{B}|\psi
\rangle _{ABC}\langle \psi |)$. The $\mathcal{N}_{A(BC)}$ is a one-tangle
and is defined as $\mathcal{N}_{A(BC)}=\left\Vert \rho
_{ABC}^{T_{A}}\right\Vert -1$, where $\left\Vert O\right\Vert =\mathrm{tr}%
\sqrt{OO^{\dag }}$ stands for the trace norm of an operator $O$ and $\rho
_{ABC}^{T_{A}}$ is the partial transposition of the density matrix over
qubit $A$.

For a bipartite density matrix $\rho _{ij,mn}$, the realignment criterion 
\cite{Rudolph,Kia} is given by $\left( \rho ^{R}\right) _{im,jn}=\rho
_{ij,mn}$. A bipartite state is separable under realignment criterion if $%
\left\Vert \rho ^{R}\right\Vert \leq 1$ and it is entangled if the quantity $%
Q=\left\Vert \rho ^{R}\right\Vert -1$ is positive. The quantity $Q$ gives a
rough estimation of the degree of entanglement of the system. For the
detection of entanglement in tripartite states, the partial realignment
(linear contraction) \cite{Horodocki} is made over any two subsystems while
leaving the third one unchanged. That is, for a tripartite state $\sigma
_{ijk,mnp}$, the realignment map over the second and third subsystems, while
leaving the first subsystem untouched, is given by $\left( \sigma
^{R}\right) _{ijn,mkp}=\rho _{ijk,mnp}$. The other possible partial
realignment on subsystems can similarly be defined. Like a bipartite state,
a tripartite state is entangled if the quantity $Q=\left\Vert \sigma
^{R}\right\Vert -1$ is positive. In the following we use the linear
contraction criterion to investigate the behavior of tripartite entanglement
of our system of Eq. (\ref{4}). We also investigate that whether the
subsystems are separable or have some degree of entanglement when one or two
observers are in accelerated frames.

\section{Dynamics of tripartite entanglement in accelerated frames}

First of all we consider the behavior of entanglement of the bipartite
subsystems by tracing over one subsystem when only Charlie is in the
accelerated frame. The density matrices of the bipartite subsystems $\rho
_{BC(AC)}=\mathrm{tr}_{A(B)}(\rho _{ABC})$ for $r_{b}=0$ are diagonal and is
given by%
\begin{equation}
\rho _{BC(AC)}=\frac{1}{2}[\cos ^{2}r_{c}|00\rangle \langle 00|+\sin
^{2}r_{c}|01\rangle \langle 01|+|11\rangle \langle 11|].  \label{7}
\end{equation}%
where $\rho _{BC(AC)}$ represents either $\rho _{BC}=\mathrm{tr}_{A}(\rho
_{ABC})$ or $\rho _{AC}=\mathrm{tr}_{B}(\rho _{ABC})$. Applying realignment
criterion leads to the following result%
\begin{equation}
\rho _{BC(AC)}^{R}=\frac{1}{2}[\cos ^{2}r_{c}|00\rangle \langle 00|+\sin
^{2}r_{c}|11\rangle \langle 00|+|11\rangle \langle 11|].  \label{7A}
\end{equation}%
From this we can find $\rho _{BC(AC)}^{R\dag }$ and then using the
definition of trace norm we can easily calculate the quantity $%
Q_{BC(AC)}=\left\Vert \rho _{BC(AC)}^{R}\right\Vert -1$ which is given by%
\begin{equation}
Q_{BC(AC)}=\frac{1}{4}(-2+\sqrt{3+\cos 4r_{c}}).  \label{8}
\end{equation}%
One can see that Eq. (\ref{8}) gives $0$ for $r_{c}=0$, which means that
there is no bipartite entanglement between the subsystems $BC(AC)$, a result
of the inertial frames for \textit{GHZ} state. Although $Q_{BC(AC)}$\
depends on the acceleration of the moving frame, however, $Q_{BC(AC)}<0$ for
all values of $r_{c}>0$. Similarly by tracing over the Charlie qubit, the
density matrix $\rho _{AB}=\mathrm{tr}_{C}(\rho _{ABC})$ for bipartite
subsystem $AB$ is independent of the Charlie acceleration and is given by $%
\rho _{AB}=1/2(|00\rangle \langle 00|+|11\rangle \langle 11|)$. From this
matrix, it is straightforward to verify that $Q_{AB}=0$, the same as the
result for any bipartite cut in inertial frames. This means that the
bipartite subsystems have no entanglement regardless of the acceleration of
the Charlie frame. The acceleration of the noninertial frame does not
generate bipartite entanglement.

Now we use the partial realignment criterion to investigate the behavior of
genuine tripartite entanglement of the system for the case when only Charlie
is in accelerated frame. Since there are three qubits, the partial
realignment can be made over any two qubits. In the case of inertial frames
the quantity $Q$ is invariant regardless of which two qubits are realigned.
To see whether it is true in the noninertial frames as well, we apply the
partial realignment criterion to all the possible cases. First we
investigate the case when Bob's qubit and Charlie's qubit are realigned.
Under this\ partial realignment the density matrix of Eq. (\ref{4}) for $%
r_{b}=0$ becomes%
\begin{eqnarray}
\rho _{ABC}^{R_{(BC)}} &=&\frac{1}{2}[\cos ^{2}r_{c}|000\rangle \langle
000|+\sin ^{2}r_{c}|011\rangle \langle 000|  \notag \\
&&+\cos r_{c}(|101\rangle \langle 001|+|010\rangle \langle 110|)+|111\rangle
\langle 111|],  \label{9}
\end{eqnarray}%
where $R_{(BC)}$ in the superscript indicates that the partial realignment
is made over Bob's qubit and Charlie's qubit. The quantity $%
Q^{BC}=\left\Vert \rho _{ABC}^{R_{(BC)}}\right\Vert -1$ can easily be
calculated, as done in the previous cases, and is given by%
\begin{equation}
Q^{BC}=\frac{1}{4}(-2+4\cos r_{c}+\sqrt{3+\cos 4r_{c}}).  \label{10}
\end{equation}%
Note that we use $BC$ in the superscript of $Q$ in order to differentiate it
from the one used above for bipartite case. Similarly, when the partial
realignment is made over Alice's qubit and Charlie's qubit, the quantity $%
Q^{AC}=Q^{BC}$ ($Q^{AC(BC)}$), which shows that the behavior of entanglement
remains unchanged against the acceleration with respect to the partial
realignment of these two bipartite subsystems. However, the result is
different when the realignment is made over the two inertial observers'
qubits, that is, on Alice's qubit and Bob's qubit. In this case the quantity 
$Q^{AB}=\cos r_{c}$. This means that the entanglement quantifier $Q$ in the
noninertial frame is dependent on which two qubits are to be partially
realigned. It can be seen that $Q^{AC(BC)}=Q^{AB}=1$ for $r_{c}=0$, which is
true for the case of inertial frames.

\begin{figure}[h]
\begin{center}
\begin{tabular}{ccc}
\vspace{-0.5cm} \includegraphics[scale=1.2]{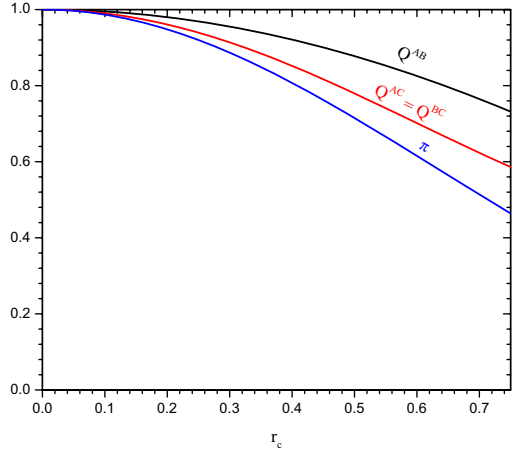}\put(-320,220) &  & 
\end{tabular}%
\end{center}
\caption{(color online) The quantities $Q^{AC(AB)}$, $Q^{AB}$ and the $%
\protect\pi $-tangle are plotted against the acceleration parameter $r_{c}$
for the case when one observer is in the accelerated frame.}
\label{Figure1}
\end{figure}

To see the effect of acceleration on the genuine tripartite entanglement
more deeply, we plot the quantities $Q^{AC(BC)}$ and $Q^{AB}$ against the
acceleration parameter $r_{c}$ in Fig. (\ref{Figure1}). It is shown that
both $Q^{AC(BC)}$ and $Q^{AB}$ decrease monotonically from its maximum value 
$1$ at zero acceleration to some minimum positive values at infinite
acceleration. The decrease in $Q^{AC(BC)}$ is faster than in $Q^{AB}$. We
think that the slower decrease in $Q^{AB}$ against $r_{c}$ is due to the
part of three way entanglement shared between inertial observers which is
unaffected directly by the acceleration of Charlie frame. Unlike the
bipartite entanglement in noninertial frames which goes to zero in the limit
of infinite acceleration, the positive values of $Q^{AC(BC)}$ and $Q^{AB}$
show that the tripartite state is entangled even at infinite acceleration.
For comparison of our results with the results of Ref. \cite{Wang2}, we have
also plotted the $\pi $-tangle of one observer accelerated case in Fig. (\ref%
{Figure1}). It shows that like for many bipartite and tripartite states in
inertial frames, for tripartite \textit{GHZ} state the quantity $Q\geq \pi $
($\pi $-tangle) in the noninertial frames.

Next we consider the case in which two observers, Bob and Charlie, are in
accelerated frames. We consider the simplest case in which $r_{b}=r_{c}=r$,
the realigned matrix for the bipartite subsystem $BC$ from Eq. (\ref{4}) in
this case becomes%
\begin{eqnarray}
\rho _{BC}^{R} &=&\frac{1}{2}[\cos ^{4}r|00\rangle \langle 00|+\cos
^{2}r\sin ^{2}r(|11\rangle \langle 00|+|00\rangle \langle 11|)  \notag \\
&&+(1+\sin ^{4}r)|11\rangle \langle 11|].  \label{11}
\end{eqnarray}%
The quantity $Q_{BC}$ can easily be found like done previously and is given
by%
\begin{equation}
Q_{BC}=\frac{1}{4}[-4+\cos ^{2}r\sqrt{3+\cos 4r}+2\sqrt{\cos ^{4}r\sin
^{4}r+(1+\sin ^{4}r)^{2}}].  \label{12}
\end{equation}%
Similarly, for the other two subsystem $Q_{AB}=Q_{AC}$ and is given by%
\begin{equation}
Q_{AB(AC)}=\frac{1}{4}(-2+\sqrt{3+\cos 4r}).  \label{13}
\end{equation}%
Note that for $r\geq 0$, $Q_{BC},Q_{AB(AC)}\leq 0$ which show that no
entanglement exists in the bipartite subsystems.

To find the genuine tripartite entanglement we proceed as before. In this
case, partially realigning the density matrix of Eq. (\ref{4}) over Bob's
qubit and Charlie's qubit, we obtain%
\begin{eqnarray}
\rho _{ABC}^{R_{(BC)}} &=&\frac{1}{2}[\cos ^{4}r|000\rangle \langle
000|+\sin ^{4}r|011\rangle \langle 011|  \notag \\
&&+\cos ^{2}r(|101\rangle \langle 001|+|010\rangle \langle 110|)  \notag \\
&&+\sin ^{2}r\cos ^{2}r(|011\rangle \langle 000|+|000\rangle \langle 011|) 
\notag \\
&&+|111\rangle \langle 111|].  \label{14}
\end{eqnarray}%
From Eq. (\ref{14}), the quantity $Q^{BC}$ can straightforwardly be found by
using the definition of trace norm. It is given by%
\begin{equation}
Q^{BC}=\frac{1}{4}(-2+4\cos ^{2}r+\sqrt{3+\cos 4r}).  \label{15}
\end{equation}%
Interestingly enough, unlike the previous case if we partially realign Eq. (%
\ref{4}) over the other two sets of qubits the behavior of tripartite
entanglement remains unchanged. In other words, the quantities $Q^{AC}$ and $%
Q^{AB}$ are both equal to $Q^{BC}$. However, for the case of different
acceleration of the two observers, this may not be true and needs to be
checked.

\begin{figure}[h]
\begin{center}
\begin{tabular}{ccc}
\vspace{-0.5cm} \includegraphics[scale=1.2]{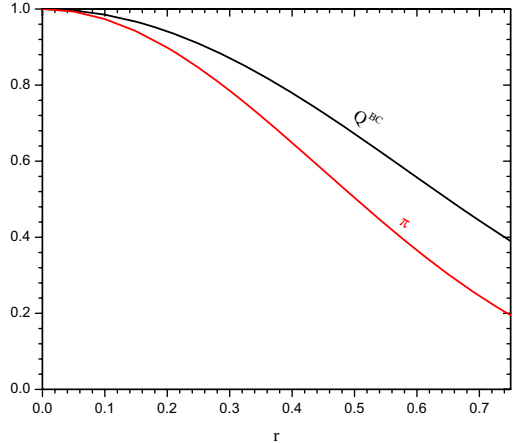}\put(-320,220) &  & 
\end{tabular}%
\end{center}
\caption{(color online) The quantity $Q^{BC}$ and the $\protect\pi $-tangle
are plotted against the acceleration parameter $r$ for the case when two
observers are in accelerated frames.}
\label{Figure2}
\end{figure}

To see the behavior of tripartite entanglement, we plot the quantity $Q^{BC}$
against the acceleration parameter $r$ in Fig. (\ref{Figure2}). The quantity 
$Q^{BC}=1$ for $r=0$. With increasing value of the acceleration, the
quantity $Q^{BC}$ decreases monotonically and reaches a minimum positive
value at infinite acceleration. As expected, the decrease in $Q^{BC}$ in
this case is faster than the decrease in $Q^{AC(BC)}$ and $Q^{AB}$ of the
one observer accelerated case (Fig. (\ref{Figure1})). Since the loss of
entanglement occurs due to the leakage of information into the inaccessible
regions \cite{Adesso,Wang3} as a result of the Unruh effect. The quicker
decrease in entanglement with the increasing acceleration arises because the
information leakage is higher due to the acceleration of two observers as
compare to the case of one accelerated observer. Again, a comparison of our
results for the two observers accelerated case with the results of Ref. \cite%
{Wang2} shows that for tripartite \textit{GHZ} state the quantity $Q\geq \pi 
$ ($\pi $-tangle) in the noninertial frames as shown in the figure. As the
two measurements, the $\pi $-tangle and the partial realignment criterion,
of tripartite entanglement agree that the tripartite entanglement in
noninertial frames does not completely vanish even in the limit of infinite
acceleration. Therefore, it is natural to expect that in noninertial frames
the tripartite entanglement may prove a better resource than the bipartite
entanglement for various quantum information tasks. For example, quantum
teleportation through tripartite entanglement might be possible even if some
observers falls into the black hole while others hover outside the event
horizon. However, further investigations through other measurements of
entanglement using different possible tripartite states may provide a better
understanding of the so far explored results.

\section{Conclusion}

The behavior of tripartite entanglement and bipartite subsystems of
fermionic \textit{GHZ} state is investigated when one or two observers are
in the accelerated frames. We used the realignment criterion, which is a
strong criterion for the detection of entanglement and easy in calculation,
for the investigation of the dynamics of the entanglement. It is shown that
both in one and in two observers accelerated cases the bipartite subsystems
have no entanglement, that is, all the entanglement of the system is in the
form of genuine tripartite entanglement. The acceleration of observers does
not generate bipartite entanglement between subsystems of the composite
system. This means that the entanglement resource cannot be utilized by any
two observers without the cooperation of the third one. In one observer
accelerated case, the quantity $Q$ that measures the entanglement depends on
which qubits are realigned. As commented above, we think that this
difference in $Q$ is due to the part of tripartite entanglement that is
shared between the inertial observers and is unaffected directly by the
acceleration of the accelerated observer. However, in two observers
accelerated case, this quantity is invariant with respect to the choice of
realigned qubits for the case when the accelerations of the accelerated
observers are equal. In accelerated frames it is the Unruh effect that
causes the leakage of information to the causally disconnected regions, as a
result the entanglement decreases with the increasing acceleration.
Logically, in the tripartite state of two accelerated observers the leakage
of information to the causally disconnected regions is supposed to be more
quick that would lead to a rapid decrease in entanglement, which was
expected to result in entanglement sudden death. But contrary to the
expected result and unlike the bipartite entanglement that asymptotically
goes to zero, the tripartite entanglement does not vanish even in the limit
of infinite acceleration for both one and two observers accelerated cases.
This behavior of tripartite entanglement in noninertial frames need to be
further explored by using different entanglement quantifiers on other
tripartite states. On the basis of the agreement between our results and the
results of Ref.\cite{Wang2} one can predict that tripartite entanglement
might be a better resource for quantum information processing in noninertial
frame.\newline

\end{document}